\documentclass[journal]{IEEEtran}
\usepackage{cite}
\usepackage{amsmath,amssymb}
\usepackage{algorithmic}
\usepackage{graphicx}
\usepackage{textcomp}

\begin{document}

\title{Translating Clinical Delineation of Diabetic Foot Ulcers into Machine Interpretable Segmentation}
\author{Connah Kendrick, Bill Cassidy, Joseph M. Pappachan, Claire O'Shea, Cornelious J. Fernandez, Elias Chacko, Koshy Jacob, Neil D. Reeves, and Moi Hoon Yap, \IEEEmembership{Senior Member, IEEE}
\thanks{Date submitted for review: 30 September 2022. We gratefully acknowledge the support of NVIDIA Corporation who provided access to GPU resources and sponsored Diabetic Foot Ulcers Grand Challenges. }
\thanks{ C. Kendrick, B. Cassidy and M.H. Yap are with the Department of Computing and Mathematics, Manchester Metropolitan University (e-mail: Connah.Kendrick@mmu.ac.uk, M.Yap@mmu.ac.uk). }
\thanks{J. Pappachan is with the Lancashire Teaching Hospitals NHS Foundation Trust (e-mail: pappachan.joseph@lthtr.nhs.uk).}
\thanks{C. O'Shea is with Waikato District Health Board (e-mail: claire.o'shea@waikatodhb.health.nz).}
\thanks{C. J. Fernandez is with United Lincolnshire Hospitals NHS Trust (e-mail: drcjfernandez@yahoo.com).}
\thanks{E. Chacko is with Jersey General Hospital (e-mail: e.chacko@health.gov.je).}
\thanks{K. Jacob is with Eastbourne District General Hospital (e-mail: drkoshyjacob@gmail.com).}
\thanks{N. D. Reeves is with the Research Centre for Musculoskeletal Science \& Sports Medicine (e-mail: n.reeves@mmu.ac.uk).}
}

\markboth{Preprint}%
{Kendrick \MakeLowercase{\textit{et al.}}: Translating Clinical Delineation of Diabetic Foot Ulcers into Machine Interpretable Segmentation, Preprint}

\maketitle

\begin{abstract}
Diabetic foot ulcer is a severe condition that requires close monitoring and management. For training machine learning methods to auto-delineate the ulcer, clinical staff must provide ground truth annotations. In this paper, we propose a new diabetic foot ulcer dataset, namely DFUC2022, the largest segmentation dataset where ulcer regions were manually delineated by clinicians. We assess whether the clinical delineations are machine interpretable by deep learning networks or if image processing refined contour should be used. By providing benchmark results using a selection of popular deep learning algorithms, we draw new insights into the limitations of DFU wound delineation and report on the associated issues. With in depth understanding and observation on baseline models, we propose a new strategy for training and modify the FCN32 VGG network to address the issues. We achieved notable improvement with a Dice score of 0.7446, when compared to the best baseline network of 0.5708 and the first place in DFUC2022 challenge leaderboard, with a Dice score of 0.7287. This paper demonstrates that image processing using refined contour as ground truth can provide better agreement with machine predicted results. Furthermore, we propose a new strategy to address the limitations of the existing training protocol. For reproducibility, all source code will be made available upon acceptance of this paper, and the dataset is available upon request.
\end{abstract}


\section{Introduction}
\label{sec:introduction}

\IEEEPARstart{D}{iabetic} Foot Ulcers (DFU) are caused when sections of the foot and skin are damaged due to multiple factors including nerve damage (diabetic peripheral neuropathy) and foot deformities. DFU healing can be impaired due to blood flow (vascular) limitations as a consequence of diabetes. Owing to this, the DFU requires regular checks to ensure optimal healing and to inform any adjustments to the treatment strategy. DFU frequently become infected, can lead to amputation and in some cases loss of life if antibiotic treatment is unsuccessful \cite{ghanassia2008long}.

It is shown that at least 10\% of people with diabetes will have some form of DFU in their lifetime, rising to 25\% depending on life-style factors \cite{jeffcoate2003diabetic, cavanagh2005treatment}. Moreover, recent studies have shown that after treatment, patients have a 70\% chance of ulcer recurrence \cite{ogurtsova2021cumulative}. Although DFU is a physical disease, it has also been widely reported to have a drastic impact on patient mental well-being and quality of life, causing anxiety and depression \cite{ahmad2018anxiety}.

Treatment for DFU can be a long-term process, due to diabetes-related complications impairing the healing process \cite{Davis2018}. It requires a multi-disciplinary team \cite{edmonds2021current} to monitor the progress of the ulcer, focusing largely on the management of diabetes \cite{sorber2021diabetic} and blood flow to the foot. However, complications, such as infection \cite{chang2021strategy} significantly prolong treatment. If treatment is prolonged, the possibility of infection and amputation increase significantly \cite{glover20213d}. This has been shown to create a heavy burden on healthcare systems, in terms of both time and cost per patient \cite{lo2021clinical, edmonds2021current}. Furthermore, this causes a great deal of concern due to the predicted rapid global rise of diabetes \cite{sun2021idf}, amplified significantly by the current pandemic \cite{pranata2021diabetes}. To address these challenges, researchers have been working towards development of methods \cite{yap2020deep, yap2021classification, yap2021evaluation, cassidy2021eval, yap2022overview} and automated systems capable of detecting and monitoring DFU \cite{reeves2021diabetes, cassidy2021cloudbased}. Improvements to automated delineation of DFU could support improved digital healthcare tools that could be used for screening and triage of DFU. Furthermore, these improvements could aid in the development of active DFU monitoring systems, to engage the healing process stage.

This paper demonstrates the processes of translating clinical delineation of DFU into machine interpretable segmentation. 
We contribute to the research progress of DFU segmentation in the following ways:

\begin{itemize}
    \item Introduce the largest DFU segmentation dataset to date with ground truth delineation (namely, DFUC2022) and perform detailed analysis.
    \item Investigate the effect of image processing refined contours on the performance of a popular deep learning segmentation algorithm, DeepLabv3+.
    \item Establish baseline results for the DFUC2022 dataset using a range of popular deep learning segmentation networks.
    \item Propose a new strategy to optimise the performance of DFU segmentation in an end-to-end network and achieved the best result when compared to the DFUC2022 challenge leaderboard's results.
\end{itemize} 

\begin{table*}[!h]
    \centering
    \renewcommand{\arraystretch}{1.0}
    \caption{A comparison of the proposed DFUC2022 datasets and the existing DFU image segmentation datasets.}
    \label{tab:datasets}
    \scalebox{1.0}{
    \begin{tabular}{|p{2.6cm}|p{0.8cm}|p{2cm}|p{1.5cm}|p{2cm}|p{1cm}|p{1cm}|}
		\hline
		Publication & Year & Dataset Name & Resolution&  Train& Test& Total\\\hline
		Wang et al. \cite{wang2020} & 2020 & AZH wound care dataset & $224\times224$& 831 & 278 &1109 \\ \hline 
		Thomas \cite{Thomas2020} & NA & Medetec & $560\times391$ \newline $224\times224$& 152 & 8 & 160\\ \hline 
		Wang et al. \cite{wang2021zenodo} & 2021 & FUSeg Challenge &$512\times512$&1010 & 200 & 1210 \\ \hline
		Proposed & 2022 & DFUC2022 &$640\times480$&2000 & 2000 & 4000 \\ \hline
		\end{tabular}
		}
	\end{table*}
	
This work will benefit the research community by providing a summary of available datasets to access and use for training segmentation based networks. With our established partnerships between clinicians and researchers, we provide the largest DFU segmentation dataset with superior image resolution when compared with existing DFU datasets \cite{wang2021zenodo}. Additionally, we provide an in-depth analysis on the performance of baseline results and propose a new end-to-end network, resulting in superior performance when compared to the best reported model in the challenge leaderboard. To assist in fair assessment and comparison with the benchmarks, we release a testing set that can be evaluated online via a grand challenge website, providing almost instant evaluation results on a standard set of performance metrics.

\section{Related Work}

\subsection{Previous Datasets} 
\subsubsection{DFUC2020 Dataset}
The DFUC2020 Dataset \cite{yap2020deep} is an object detection based dataset, containing 2000 training, 200 validation and 2000 testing images. All images are $640 \times 480$, but some images contained multiple DFUs, increasing the total number of detection annotations. Three cameras were used for image capture, Kodak DX4530, Nikon D3300 and Nikon COOLPIX P100. The images were acquired with close-ups of the full foot at a distance of around 30–40 cm with the parallel orientation to the ulcer. The use of flash was avoided, and instead, room lights were used to provide consistent colours in the images. Images were acquired by a podiatrist and a consultant physician with specialization in the diabetic foot, both with more than 5 years professional experience. All images were captured without the use of a tripod. 

\subsubsection{DFUC2021 Dataset}
The DFUC2021 dataset \cite{yap2021classification} is a multi-class DFU dataset, targeting DFU, infection, Ischaemia and both. The dataset contains 5,955 training images, and 5,734 for testing. Additionally, 3,994 images were released unlabeled to support semi and self-supervised methods. Images were captured under the same setting as the DFUC2020 dataset. 

\subsubsection{FUSeg dataset}
Wang et al. \cite{wang2022fuseg} introduced the Foot Ulcer Segmentation Dataset. This work focused on the development of segmentation CNNs using 1210 foot photographs exhibiting DFU which were collected over a 2 year period from 889 patients. They provided ground truth masks provided by wound care experts. However, many of the images were heavily padded to standardise image dimensions for training purposes. Additionally, although the images were shared as lossless PNG files, they exhibited notable compression artefacts, indicating that the original images had been heavily compressed before being converted to PNG. The provided ground truth files also appeared to be a mix of human and machine-generated masks. The images were $512\times 512$ with 1000 for training and 200 for test. The capture equipment was a Canon SX 620 hs and an iPad Pro. The AZH wound care and Medetec datasets, see Table \ref{tab:datasets} , where both used as part of the FUSeg dataset. It is noted that the AZH dataset is cropped to the ulcer region, where as the final images in the FUSeg challenge have surrounding regions.

\subsection{Related Methods}
The first works in DFU segmentation using fully convolutional techniques were completed by Goyal et al. \cite{goyal2017fully}. They performed segmentation experiments using a small dataset comprising 705 images with an FCN-16s network. They used 5-fold cross-validation with two-tier transfer learning resulting in a Dice Similarity Coefficient of 0.794 (±0.104) for segmentation of DFU region. These results were promising, however, the small size of the dataset is likely to impact the model's ability to generalise in real-world use.

More recently, the winning team of the FUSeg challenge, Mahbod et al. \cite{mahbod2021ensemble}, used an ensemble of LinkNet and U-Net networks. They achieved a Dice Similarity Coefficient of 0.888. They used pretrained weights (EfficientNetB1 for LinkNet and EfficientNetB2 for U-Net) with additional pretraining using the Medetec dataset. The challenge concluded that segmentation of small isolated areas of the wound with ambiguous boundaries were the most challenging aspects of the task. Conversely, segmentation of relatively larger wound regions showing clear boundaries where wound beds were cleansed, removing dead tissue, provided superior results. Cases clearly exhibiting infection, slough, or other impediments were also shown to provide improved results.

Current works in DFU segmentation show promising results. However, there are notable limitations to the datasets that were used to train these models. Aspects such as the quality and number of images may present issues that would negatively affect real-world application.

\begin{figure*}[!h]
	\centering
	\begin{tabular}{ccc}
		\includegraphics[width=5.42cm,height=4.9cm]{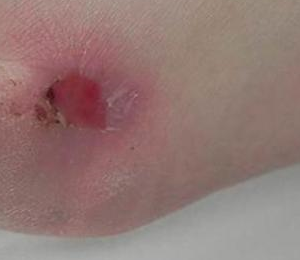} &
		\includegraphics[width=5.42cm,height=4.9cm]{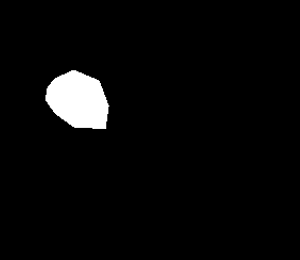} &
		\includegraphics[width=5.42cm,height=4.9cm]{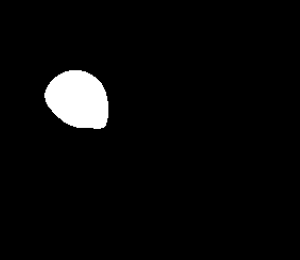} \\
		(a) & (b) & (c) \\  
	\end{tabular}
	\caption[]{Illustration of (a) an early onset DFU; (b) expert delineation; and (c) refined active contour shape.}
	\label{fig:MaskImageSnake}
\end{figure*}

\section{The DFUC2022 Dataset}
We have received approval from the UK National Health Service Research Ethics Committee (reference number is 15/NW/0539) to use DFU images for the purpose of this research. This paper introduces the largest DFU segmentation dataset, which consists of a training set of 2000 images and a testing set of 2000 images. 

 \subsection{Dataset Construction}
 The dataset was constructed in collaboration with the medical experts from Lancashire Teaching Hospitals, Waikato District Health Board, United Lincolnshire Hospitals, Jersey General Hospital, and Eastbourne District General Hospital. The DFUs were capture at room lighting, in full foot view, around 30-40cm away with the DFU centered. There cameras were used, i.e., Kodak DX4530, Nikon D3300 and Nikon COOLPIX P100. All images were taken by experienced podiatrist and physician in foot clinic. Images were then downsampled to $640\times480$ and stored in JPG format.
 
 \subsection{Reference Annotation Protocol}
 The ulcer regions on these images were delineated by experienced podiatrists and consultants. The podiatrists used the VGG annotator software, to produce a polygon outline of the DFU region in JSON format. The JSON files were then converted into binary mask images and stored in PNG format. We then preprocess the raw masks with an active contour algorithm \cite{kroon2022snake}.

 Figure \ref{fig:MaskImageSnake} illustrates an example of a DFU image showing a preprocessed region with active contour together with the expert delineation. Note that the boundary of the region is smoother after the preprocessing stage. To ensure that this smoothing process does not alter the clinical delineation, we report the agreement between expert delineation and refined contours, which produced a high agreement rate with a Dice Score of 0.9650 $\pm$ 0.0226 and Mean Intersection Over Union (mIoU) of 0.9332 $\pm$ 0.0408. These metrics demonstrate that preprocessing did not significantly alter clinical delineation, where the number of DFUs are equivalent before and after preprocessing. 
 
\begin{figure*}[!h]
\centering
\includegraphics[width=0.95\textwidth]{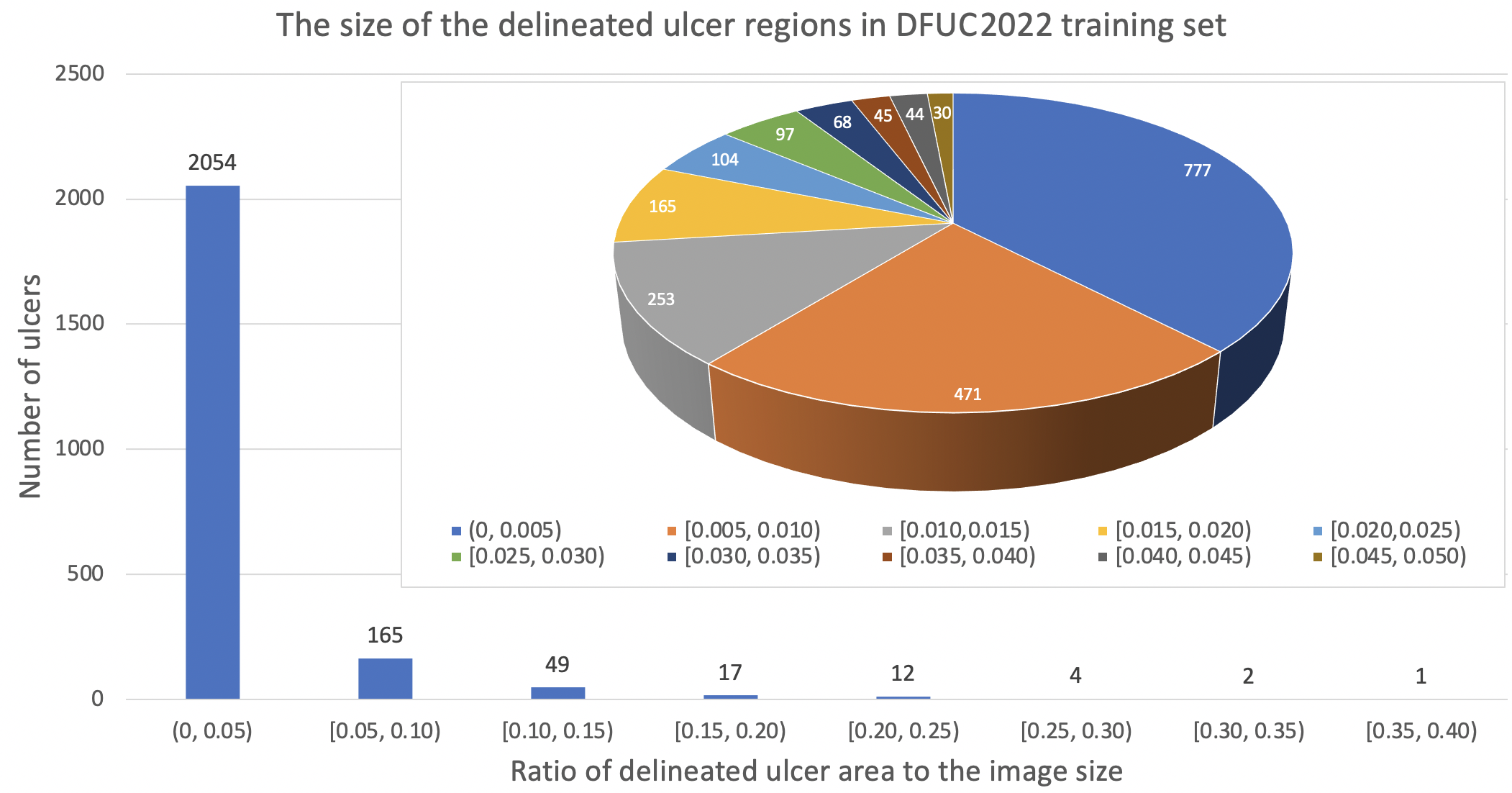}
\caption{The size distribution of delineated ulcer regions in the DFUC2022 training set. It is noted that the majority of the ulcers are smaller than 5\% of the image size.}
\label{figure: trainset}
\end{figure*}

 The DFUC2022 training set consists of 2304 ulcers, where the smallest ulcer size is 0.04\% of the total image size, and the largest ulcer size is 35.04\% of the total image size. Figure \ref{figure: trainset} provides an overview of the ratio of the delineated ulcer region to the total image size, where 89\% (2054 out of 2304) of the ulcers are less than 5\% of the total image size. The smaller images in particular represent a significant challenge for segmentation algorithms as it is widely known that deep learning algorithms have a tendency to miss small regions \cite{Goyal2017}. Another advantage of our dataset is that of the 2000 training images, there are 2304 ulcers with an average of 1.152 ulcers per image.

\section{Methods}
This section describes the methods used to investigate the effect of image processing refined contours, summarises a range of popular baseline methods for medical image segmentation, and a new strategy to improve the performance of the best segmentation method on the DFUC2022 dataset. We provide segmentation masks for the training set only, and use the grand-challenge website (https://dfuc2022.grand-challenge.org/) to allow researchers to test their methods on an exclusive testing set. We provide a total of 4000 images with 2000 binary masks for training. The masks are coded 0 for background and 1 for the DFU region.

\subsection{Manual delineation vs refined contours}
While deep learning has gained popularity in biomedical image segmentation, there are unanswered questions concerning ground truth annotation, such as: (1) would deep learning algorithms learn better with expert manual delineations (polygonal outlines) or image processing refined contours; and (2) which contour should be used for machine learning algorithms? To answer these questions in the context of DFUC2022, we run experiments with Deeplabv3+ \cite{chen2017deeplab}, one of the popular deep learning algorithms for medical imaging research \cite{khan2020prostate, azad2020skin}. Our intention is not to produce the best result, but to study the effect of coarse and detailed delineation on deep learning algorithms. Therefore, we select this algorithm without bias. First, we train two models using the default setting of Deeplabv3+, one on expert delineation and another on refined contour. We split the 2000 training images into 1800 images as training set and 200 images as validation set. Then, we test each model on the 2000 test set by using both expert delineation and refined contour as ground truth.

\subsection{Baseline methods}
We implement a wide range of existing deep learning segmentation models for the DFUC2022 baseline. These models cover a range of segmentation architectures, namely FCN \cite{long2015fully}, U-Net \cite{ronneberger2015u} and SegNet \cite{badrinarayanan2017segnet} with varying backbones to process the data, such as VGG \cite{simonyan2014very} and ResNet50 \cite{he2016deep}. We also include a comparison of alternative network depths. The range of model diversity aims to provide a good indication of techniques suitable for DFU segmentation. These new insights can direct future works with a baseline to compare against and reduce the need for repeat training of these networks. In addition to the standard U-Net, SegNet models, we provide baselines for FCN8, FCN32, U-Net and SegNet with ResNet50 and VGG as backbones.

For training the baseline networks, we use all 2000 training images, with 200 separate images for validation. We train the networks with the AdaDelta optimizer and a suggested learning rate of $0.001$, decay of $0.95$, a stabilisation epsilon of $1e-07$ as illustrated in Equation \ref{eq:adadelta1}, and using categorical cross-entropy loss, as in Equation \ref{eq:CE}. 

\begin{equation}
    E[g^{2}]_{t}=\rho E[g^{2}]_{t-1}+(1-\rho )g_{t}^{2} 
    \label{eq:adadelta1}
\end{equation}

\begin{equation}
    CE = -\sum_{i=1}^{o}Y_i\cdot log\hat{X}_i
    \label{eq:CE}
\end{equation}

\noindent
where $Y_i$ is the $i$-th ground truth value and $\hat{X}_i$ is the predicted value a $i$. We train on multiple batch sizes (2, 32, and 96), Equation \ref{eq:adadelta2} and report the best result, as defined by \cite{Zeiler2012}. 
\begin{equation}
     \Delta w_{t}=-\frac{\eta }{\sqrt{E[g^{2}]_{t}+\varepsilon }}
     \label{eq:adadelta2}
\end{equation}
\noindent
We do not perform augmentation during training or post-processing on the final prediction masks, as our aim is to produce baselines and understanding of the DFUC2022 dataset. We train the networks until the validation accuracy fails to improve, with a patience of 10 epochs.

\begin{table*}[!h]
	\centering
	\renewcommand{\arraystretch}{1.2}
	\caption{Comparison of the results when training with manual delineation as ground truth vs image processing refined contour as ground truth. The results show the machine predicted masks have better agreement with the refined contour.}
	\scalebox{1.2}{
	    \label{tab:whichco}
		\begin{tabular}{|c|c|cc|}
			\hline
			Train & Test & \multicolumn{2}{c|}{Metrics}\\
			& & Dice & mIoU \\
			\hline
			\hline
			Manual delineation & Manual delineation & 0.5870$\pm$0.3135 & 0.4809$\pm$0.2993\\ 
			Manual delineation & Refined contour & 0.5930$\pm$0.3131 & 0.4871$\pm$0.2999 \\ 
			Refined contour &Manual delineation & 0.6219$\pm$0.0286 & 0.5162$\pm$0.2967  \\ 
			Refined contour &Refined contour & \textbf{0.6277$\pm$0.3051} & \textbf{0.5224$\pm$0.2967}\\
			\hline
		\end{tabular}
    }
\end{table*}

\subsection{Challenge Competition}

To enable open research the DFUC2022 dataset was released in three parts between the 27th April 2022 and the 1st July 20222:
\begin{itemize}
    \item Training dataset, 2000 images: 27th April 2022.
    \item Validation dataset, 200 images: 21st June 2022.
    \item Test dataset, 2000 images: 1st July 2022.
\end{itemize}

At the release of the validation and test dataset, we released online submissions for live testing. We closed the online submissions on the 29th July 2022, during this time participants could analyse their methods via the validation scores. After the release of test results, we opened a live testing leaderboard to allow future submissions. We compare against the top-10 results in the challenge leaderboard.

\subsection{Proposed method}
Results from the baseline models highlighted a number of issues, such as pixelation and a high number of false positives (small regions). 
\begin{figure*}
	\centering
	\includegraphics[scale=0.44]{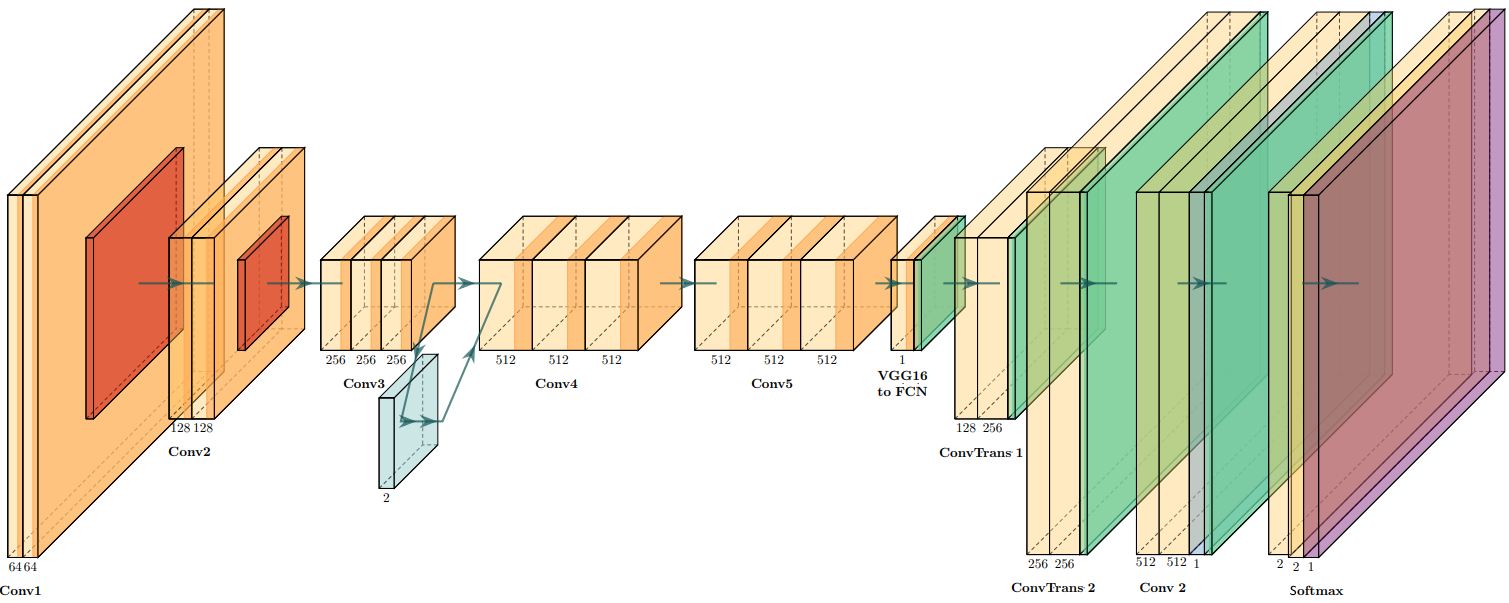}
    	\caption{Illustration of the network structure. Orange represents convolutional layers with Leaky ReLU activation, red indicates max pooling, and light green indicates skip connections using modified squeeze and excite. In the decoding section, green is a dropout layer, yellow is a separable convolution, with dilation, and the softmax layer.}
	\label{fig:network}
\end{figure*}
Previous research uses post-processing methods to improve performance. Instead of using morphology, we propose a new strategy using an modified end-to-end deep learning network to enable improved learning of our dataset, and remove the post-processing process. We use the FCN32 architecture with VGG as backbone, as shown in Figure \ref{fig:network}. First, we replace the standard ReLU layer in the full network with Leaky-ReLU, depicted by Equation \ref{eq:leakyReLU}.
\begin{equation}
    \mathit{f}\left ( x \right ) =\left\{\begin{matrix}\alpha x \, for \, x < 0
 \\
x \, for \, x \geq 0
\end{matrix}\right.
    \label{eq:leakyReLU}
\end{equation}
\noindent
where $\alpha$ is a scalar for sub zero values and $x$ the input, with an alpha of $0.3$ which aids network learning as it prevents dead neurons occurring. Then, we target excessive down-sampling by removing the bottom three max-pooling layers, while maintaining the padding. This process retains the feature map size from the lowest in the standard network of $20 \times 15$ to $160 \times 120$ on the full size images, improving the ability of the network to maintain feature maps of smaller ulcers and tracks overall wound shape, which reduces the issues with biases in dataset distribution. 

To resolve the issue of background noise, we experimented using gated convolutions \cite{dauphin2017language}. During this stage the best performing method was modified using a squeeze and excite layer \cite{hu2018squeeze} after the final pooling was used, where a dilated convolution (kernel size $5 \time 5$, dilation rate $2 \times 2$) focused on separating the foot region features from the background, which had a standard convolution (kernel size $1 \times 1$) with sigmoid activation. The resulting feature map was multiplied against the normal output of the 3rd pooling stage of the network. These adjustments resulted in improved removal of noisy inconsistent data, reducing the background features of the environment and improving focus on the more consistent foot regions. Thus, the lower levels of the networks can separate the similar textural features of the DFU and foot region. We then address the issue of rapid up-sampling by adjusting the FCN network to gradually grow the predictions through a series of small transposed convolutions (kernel size $2 \times 2$, stride $2 \times 2$) with a convolution to refine the contours of the up-sample until the desired size is reached. In many segmentation tasks, post processing of outputs is performed to smooth predictions and blob removal, however we accomplish this internally within the network with a final dilated convolution, as shown in Equation \ref{eq:dilatedConv} (kernel size $3 \times 3$, dilation rate $2$). 

\begin{equation}
    (F*_{l} k)(p) = \sum\limits_{{s + lt = p}} F (s)K(t)
    \label{eq:dilatedConv}
\end{equation}
\noindent
where $_l$ is the dilation rate providing a gap between receptive points. $K(t)$ is the values of the filter. $F (s)$ is the input to the layer and $\sum\limits_{{s + lt = p}}$ is the sum of the receptive fields. This allows for the surrounding regions to determine if the section is a small island for removal, or an edge for smoothing, using the wider receptive field.

\begin{figure*}[!h]
    \centering
	\includegraphics[scale=0.75]{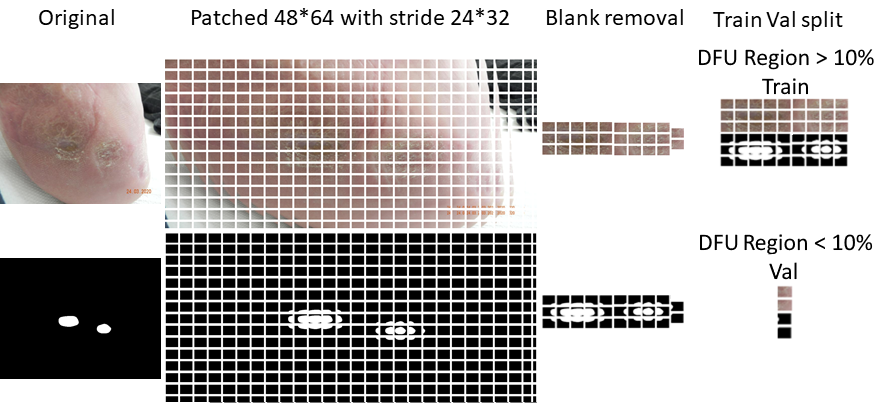}
    	\caption{Illustration of the patching system used for creating the training and validation sets. We use a half stride to create the image windows, to increase the dataset size and reduce chances of only edge cases. We then remove all blank patches from the set, use all with greater then 10\% DFU pixels for training and any other for validation.}
	\label{fig:TrainProcessing}
\end{figure*}

\begin{figure*}
    \centering
	\includegraphics[scale=0.55]{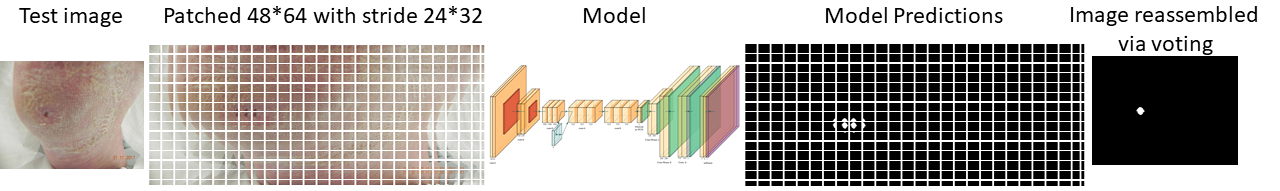}
    	\caption{Illustration of the Testing pipeline. We test on all patches of the images without removal to ensure the system is capable of predicting under a wide array of backgrounds. We then use a vote when reconstructing the image, due to the stride overlap where all must agree to be classed as an ulcer.}
	\label{fig:TestProcessing}
\end{figure*}
We also adjusted the training routine using a weighted loss function, which showed further improvements. However, for better results we used the standard loss function and fully balanced the dataset, we processed the training dataset to crop out sliding windows of $64 \times 48$ with a stride of $32 \times 24$, as illustrated in Figure \ref{fig:TrainProcessing}. The stride allowed the network to obtain as much of the wound features as possible, producing a total of 810,000 patches. Next, all the patches from the set that contained no DFU pixels were removed, leaving 55,760 patches with DFU pixels. After this, we processed the images to create the training and validation sets, by moving any images with less than 10\% DFU pixels into the validation set and using all others for training, giving a total of 38,997 patches for training and 16,763 patches for validation. This stage provided two key advantages:

\begin{itemize}
    \item \textit{Balanced split of classes}: In total the amount of background pixels was 51.71\% and DFU pixels was 48.29\%. Thus, giving a more balanced set compared to the standard training method, meaning that both classes will have even weighting. 
    \item \textit{Difficult validation set}: The validation set was heavily biased towards background features. Many of the validation case were small edge cases which are particularly challenging for segmentation networks. This means that a good score reflects a network with clear data understanding.
\end{itemize}

For the modified network, we train on a batch size of 2, providing the network a balanced view of the data. The same settings for optimizer, learning rate, and loss function are used as in the baseline methods. The network was adjusted to take in the patches are their current resolution. For the modified network, the test dataset was also split in the same process of $64 \times 48$ with a stride of $32 \times 24$. To reconstruct the image overlapping sections, due to the stride, all patches had to agree for the pixel to be classified as ulcer, as show in Figure \ref{fig:TestProcessing}.

\begin{table*}[!h]
	\centering
	\renewcommand{\arraystretch}{1.2}
	\caption{A comparison of the overall performance of the state-of-the-art methods with and without pretrained model, results reported on their best epoch. $\dagger$ = higher score is better; $\uplus$ = lower score is better. We train 12 unique models with different batch sizes. However, we only show the models with the settings that resulted in the best performance. \textit{Italic} indicates the best baseline result and \textbf{bold} indicates the best overall result.}
	\scalebox{1.2}{
	    \label{table:baselines}
		\begin{tabular}{|l|l|c|cccc|}
			\hline 
			Model & Backbone  & Settings & \multicolumn{4}{|c|}{Metrics} \\
			
		 &&Best Batch Size & Dice $\dagger$ & mIoU $\dagger$ & FPE $\uplus$ & FNE $\uplus$  \\
			\hline
			\hline
FCN8 &&2&0.2621&0.1914&0.6789&0.6062\\

&ResNet50&2&0.4993&0.3963&0.4576&\textit{0.3824}\\

&VGG&2&0.5101&0.3952&0.3643&0.4500\\
			\hline
FCN32&&2&0.2174&0.1594&0.7564&0.6980\\

&ResNet50&2&0.4334&0.3372&0.5090&0.5081\\

&VGG&2&\textit{0.5708}& \textit{0.4549}& 0.3396&0.3833\\
\hline
SegNet& &32&0.2677&0.1880&0.6325&0.6510\\

 &ResNet50&32&0.4768&0.3676&0.4325&0.4339\\

& VGG &32&0.4596&0.3469&0.4003&0.5158\\
\hline
U-Net&&2&0.4057&0.3035&0.4900&0.5119\\

&ResNet50&32&0.0646&0.0371&0.5585&0.9584\\

&VGG&2&0.1446&0.0878&\textit{0.2501}&0.9020\\
\hline \hline
Proposed&Modified VGG &2 & \textbf{0.7447} & \textbf{0.6467} & \textbf{0.1866} & \textbf{0.3056}\\ 
\hline
		\end{tabular}
    }
\end{table*}

\subsection{Performance metrics}
In image segmentation, the commonly used evaluation metrics are: \\
Dice Similarity Index as shown in Equation (\ref{eq:DICE}):
\begin{equation}
   Dice = 2 * \frac{|X \cap  Y|}{|X| + |Y| - |X \cap Y|}
    \label{eq:DICE}
\end{equation}

\noindent
 and Intersection Over Union (IoU) (also known as Jaccard Index) as shown in Equation (\ref{eq:JACCARD}):  
    \begin{equation}
       IoU = \frac{|X \cap  Y|}{|X| + |Y|}
        \label{eq:JACCARD}
    \end{equation}
\noindent
where $X$ and $Y$ represent the ground truth mask and the predicted mask. We used mIoU to better represent the segmentation outcomes for both classes (ulcer and background).

We include additional metrics to understand Type I and Type II errors of the algorithm performance. These two additional metrics are: \\
False Positive Error (FPE) as in Equation (\ref{eq:FPE}):
\begin{equation}
  FPE = \frac{FP}{FP + TN}
    \label{eq:FPE}
\end{equation}

\noindent
 and False Negative Error (FNE) as in Equation (\ref{eq:FNE}).

\begin{equation}
   FNE = \frac{FN}{FN + TP}
    \label{eq:FNE}
\end{equation}

\noindent
where $FP$ is the total number of false positives in the predictions, $TN$ is the total number of true negatives and $FN$ is the total number of false negatives. 

\begin{figure*}[!h]
	\centering
	\includegraphics[scale=0.72]{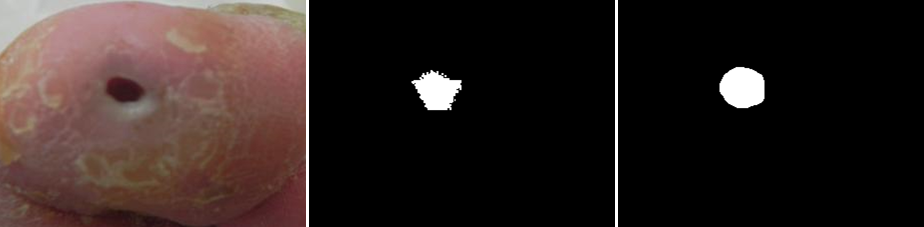}
    	\caption{An example of how the inclusion of dilation smoothing improves predictions in the modified network on full images. From left to right: input image, standard FCN32 VGG and modified FCN32 VGG. \textit{Note: For illustration, images were cropped to focus on DFU region.}}
	\label{fig:smooth}
\end{figure*}

\section{Results}
Table \ref{tab:whichco} shows the results when trained on two types of annotation: manual delineations vs refined contours. The results show that the algorithm did not learn as effectively from the human delineation on the boundary (polygonal outlines). The refined contour consistently demonstrated closer agreement with the machine predictions, without relying on the type of ground truth used for training. Therefore, we use image processing refined contour as ground truth for both train set and test set, for the rest of the paper.


\begin{figure*}[!h]
	\centering
	\includegraphics[scale=0.55]{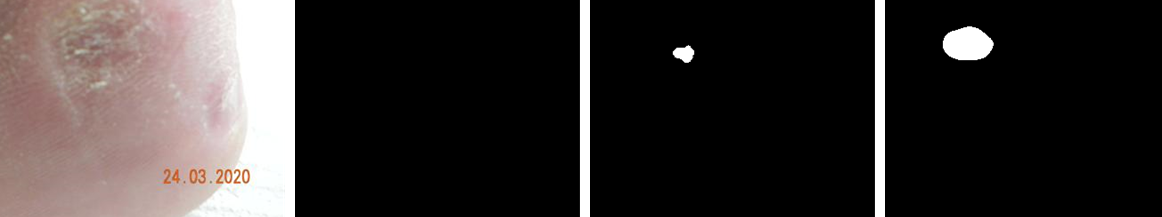}
    	\caption{An example of texture similarity and over down-sampling issues in a DFU prediction. From left to right: original image, standard FCN32 VGG, modified FCN32 VGG and ground truth. \textit{Note: For illustration, images were cropped to focus on DFU region.}}
	\label{fig:Texture}
\end{figure*} 

As shown in Table \ref{table:baselines}, many of the available techniques give reasonable results in DFU segmentation. Among the baseline methods, the best performing model was FCN32 with a VGG backbone, with the highest Dice score of 0.5708 and 0.4549 for mIoU. A key factor in this task is the ability of the network to handle images without positive DFU cases (True Negatives), thus we use the FPE metric. In such cases the best performing model is also FCN32 VGG, which shows a high understanding of the surround regions. We observe that most methods that use a higher batch size resulted in significant performance degradation. A contributing factor to this is likely to be background noise present in the images where the environment can vary significantly between images. Lower batch sizes allowed the system to focus on a case by case basis, allowing the network to slowly learn to ignore the background noise and focus on the wounds.

\begin{figure*}[!h]
    \centering
	\includegraphics[scale=0.3]{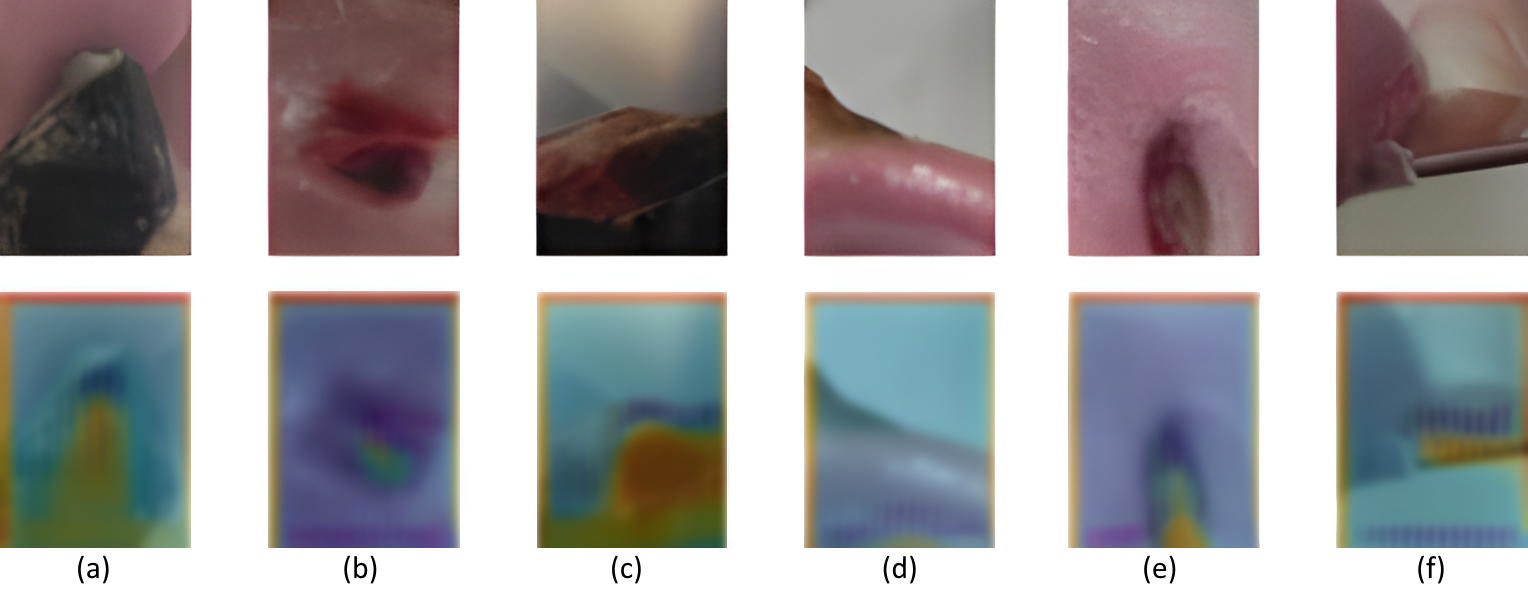}
    	\caption{Examples of the patches used in the modified network, demonstrating the ability of the network to focus on DFU regions, including edge cases (d) and occlusion (f).}
	\label{fig:Heatmap1}
\end{figure*}

\begin{figure*}[!h]
    \centering
	\includegraphics[scale=0.9]{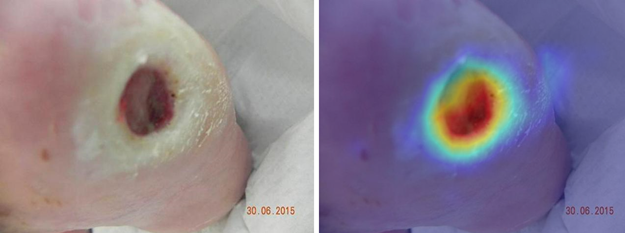}
    	\caption{Illustration of a prediction in a full resolution image, note that the network correctly focuses on DFU regions.}
	\label{fig:Heatmap2}
\end{figure*}

\begin{table}[!htb]
\centering
\renewcommand{\arraystretch}{1.0}
\caption{Comparison of the DFUC2022 entries and our proposed method.}
     \label{tab:GCscores}
 	 \begin{tabular}{|c|cccc|}
 		\hline
 		Team &  \multicolumn{4}{c|}{Metrics}\\
 		& Dice $\dagger$ & mIoU $\dagger$ & FPE $\uplus$ & FNE $\uplus$  \\
 		\hline
 		\hline
 		yllab & \textbf{0.7287} & 0.6252 & 0.2048 & 0.2341\\ 
 		LkRobotAI Lab & 0.7280 & \textbf{0.6276} & 0.2154 & 0.2261\\
 		agaldran & 0.7263 & 0.6273 & 0.2262 & \textbf{0.2210}\\
 		ADAR-LAB & 0.7254 & 0.6245 & \textbf{0.1847} & 0.2582\\
 		seoyoung & 0.7220 & 0.6208 & 0.1925 & 0.2584\\
 		FHDO & 0.7169 & 0.6130 & 0.2145 & 0.2453\\
 		GP\_2022 & 0.6986 & 0.5921 & 0.2065 & 0.2778\\
 		DGUT-XP & 0.6984 & 0.5945 & 0.2523 & 0.2379\\
 		IISlab & 0.6975 & 0.5926 & 0.2163 & 0.2734\\
 		AGH\_MVG & 0.6725 & 0.5690 & 0.2555 & 0.2830\\
 		\hline
 		\hline
 		Ours & \textbf{0.7447} & \textbf{0.6467} & 0.1866 & 0.3056\\ 
\hline
 	\end{tabular}
\end{table}

Table \ref{tab:GCscores} highlights the results for DFUC2022. The top 10 scores demonstrate the challenge of DFU segmentation for a wide range of networks. The team yllab achieved the best score in Dice (0.7287) in which the challenge was based. This was closely followed by LKRobotAI Lab, who achieved the highest mIoU (0.6276) showing a high agreement of prediction and ground truth overlap. The 3rd place team, agaldran, achieved the lowest FNE (0.2210), highlighting that they reduced the amount of falsely predicted DFU pixels, whereas the 4th place team, ADAR-LAB, achieved the best FPE score (0.1847). Our method achieves higher Dice (0.7447) and mIoU (0.6467) scores, showing a high degree of agreement between prediction and ground truth. Additionally, we have a slightly higher FPE (0.1866) when compared to the best performing (0.1847). However, one outlier with our method is that we report lowest performance in FNE (0.3056). Our method demonstrates that training using patched images and reduced pooling allows for improvements in Dice and mIoU.

The results in Table \ref{table:baselines} and \ref{tab:GCscores} show that our proposed strategy and modified network has improved the results and achieved 0.7447 for Dice, 0.6467 for mIoU, 0.1866 for FPE and 0.3056 for FNE. As visually illustrated in Figure \ref{fig:smooth}, the modified network, with the inclusion of dilation smoothing, is able to refine the results within the network. Another example to show the superiority of the modified network is in Figure \ref{fig:Texture}, where due to the similarity between surrounding skin and DFU the standard method fails, but the modified network is able to detect some overlap.

As shown in Figures \ref{fig:Heatmap1} and \ref{fig:Heatmap2}, the best performing network successfully highlights and focuses on the DFU regions. In addition, these figures highlight how the network modifications allow the system to identify a wide variety of DFU features within an image. However, note that in full size predictions (see Figure \ref{fig:Heatmap1}) the small mark to the left and the damaged skin on the right have also been focused on by the network. This highlights some of the features of the modified network segmentation, similar to Figure \ref{fig:Texture}, the broken skin could indicate a early onset of DFU. Similarly the minor activation on the left could be an indication of a smaller ulcer, due to its colour, shape and texture. Thus, a slight activation over these regions is shown.

\section{Discussion}

\begin{figure*}[!h]
    \centering
	\includegraphics[scale=0.58]{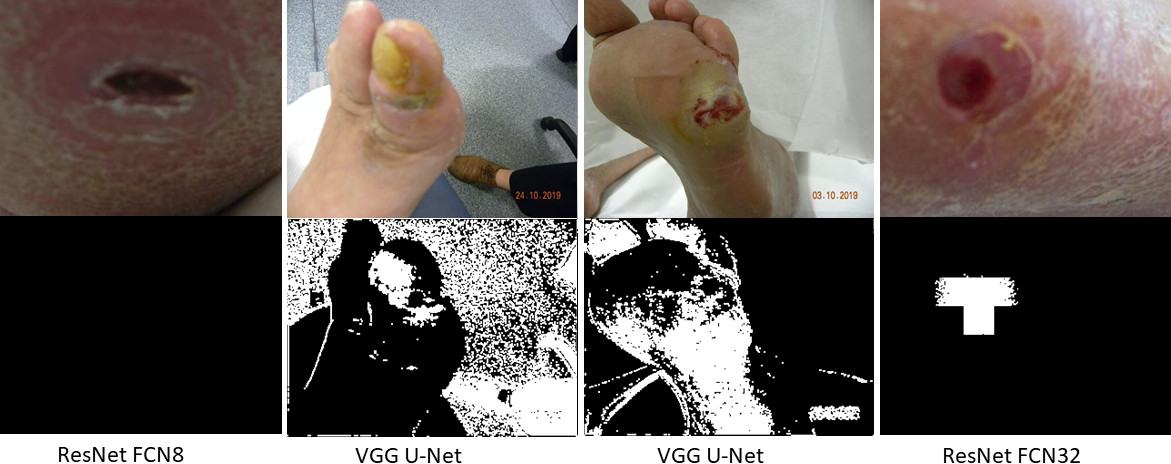}
    	\caption{ Illustration of issues associated with the baseline models: (Left) An example over down-sampling removing lesion; (Middle-left) example of background noise effecting prediction; (Middle-right) example of region similarity preventing accurate segmentation; and (Right) example of rapid up-sampling producing a block artefact. \textit{Note: some images were cropped to focus on DFU region.}}
	\label{fig:Issues}
\end{figure*}

We highlight that the best performing baseline methods had several difficulties which reduced model performance, as shown in Figure \ref{fig:Issues}: 

\begin{itemize}
    \item \textit{Excessive down-sampling of images}: Many of the segmentation backbones are based on classification networks in which reducing to core features is essential. However, with the small image to wound size ratio, this removes the full wound from the image.
    \item \textit{Data distribution}: As show in Figure \ref{figure: trainset}, a large proportion of the dataset has a DFU to background ratio of $<$5\%. This represents a large dataset bias towards none-DFU regions. This causes the networks to prioritize on the background class over DFU region, and in some cases the DFU class is ignored.
    \item \textit{Background noise}: Owing to the shape and location of DFU and patient mobility, many of the images contain a wide assortment of noise. In some cases, the foot is surrounded by a blue or white cloth so the network can focus, but in many cases the background contains clothes, floor details and other medical equipment. This poses a difficulty and the network must learn to cope with a large variety of background data. 
    \item \textit{Region similarity}: With many cases of DFU the textural quality of the lesion is similar to that of the surrounding skin, especially in cases of infection. The textural similarity of DFU regions, periwound and surrounding skin regions, introduces difficulty in distinguishing the regions, as shown in Figure \ref{fig:Texture}. This means that the networks struggled to differentiate between the DFU and other parts of the foot.
    \item \textit{Rapid up-sampling}: Due to the focus of the backbones ability to output valid feature maps the head of the network is usually light weight. This results in the up-sampling output being performed at a high rate, causing pixelated regions, in addition to small false detection regions.
\end{itemize}

These issues are the cause of the difficulties the baseline models produce. Oversampling removes the smaller wounds, which amplifies the problem of data distribution, where most wounds are below 5\% of the total image size, meaning the networks focus more on the background than on the DFU regions. Furthermore, this focus on the background data is amplified by the inconstant and noisy data. Owing to this, the region similarity of the DFU, periwound and surrounding skin is made difficult causing some networks to focus on the entire foot over the DFU regions, as there is too much focus on background data. Finally, the networks perform well using the smoothed masks over the original jagged contours provided by clinicians. Thus, in the final stages when re-upsampling to the desired size, pixelation occurs due to the rapid up-sampling, producing block-like segmentation that requires additional post-processing to smooth and remove small regions. 

\section{Conclusion}
In this paper, we introduce the largest available DFU dataset containing 2000 annotated training images and 2000 test images without annotations, together with the capability of online evaluation of network predictions. We also provide challenging cases, such as non-DFU cases and images resulting from annotator disagreement. We then provide a series of baselines on state-of-the-art models with explainable AI techniques.  

We demonstrate that by performing preprocessing on the expert delineation to smooth the DFU regions, the networks were able to produce more accurate DFU segmentation results. This was shown by comparing a cross validation between raw and smoothed masks. From this study we perform an ablation study on widely used semantic segmentation networks, producing a set of baseline results. The prediction results from the trained models highlight the difficulty in DFU clinical delineation where inter reliability can be inconsistent. This work sheds light on the challenges inherent in the development of AI systems which can aid the standardisation of DFU delineation over time to track healing progress.

We identify the shortcomings inherent in traditional segmentation networks and training techniques using the DFUC2022 dataset. From these findings we modified the best performing network and tailor it to the unique challenges presented by the DFU2022 dataset. From these adjustments to the network design we show a significant increase in model performance, without the use of post processing techniques.

Finally, we analyse heatmaps of successfully trained DFU model predictions on DFU regions, which indicate that the network is capable of focusing on ulcer regions and corresponding features when generating final prediction masks. These machine learning advancements will contribute towards supporting healthcare systems to better manage the increasing demands of DFU care, including the accurate and regular monitoring of DFU healing to increase flexibility in treatment plans.

\bibliographystyle{IEEEtran}
\bibliography{arxiv}

\end{document}